\documentclass[aps,prb,reprint,superscriptaddress,longbibliography,sort&compress]{revtex4-2}
\PassOptionsToPackage{dvipsnames,svgnames,table}{xcolor}

\usepackage{amsmath,amssymb,amsfonts,bm}
\usepackage{hyperref}
\usepackage[utf8]{inputenc}
\usepackage{graphicx}
\usepackage{dcolumn}
\usepackage{enumerate}
\usepackage{braket}
\usepackage{color,soul}
\usepackage{wasysym}
\usepackage{mathrsfs}
\usepackage{times}
\usepackage[dvipsnames,svgnames,table]{xcolor}
\usepackage{fancyhdr}
\usepackage{float}
\usepackage[english]{babel}
\usepackage[caption=false]{subfig}
\usepackage{physics}
\usepackage{mathtools}
\usepackage{upgreek}
\usepackage[normalem]{ulem}

\hypersetup{
pdfstartview={FitH},
colorlinks=false,
pdfborder={0 0 0},
linkcolor=NavyBlue,
citecolor=Maroon,
filecolor=NavyBlue,
urlcolor=NavyBlue
}

\newcommand{\bmat}{\begin{pmatrix}}
\newcommand{\emat}{\end{pmatrix}}

\begin{document}

\title{Critical timescales for chiral state conversion and irreversibility}

\author{Giorgos Pappas}
\email{Georgios.Pappas@univ-lemans.fr
}
\affiliation{Laboratoire d'Acoustique de l'Universit\'e du Mans (LAUM), UMR 6613, Institut d'Acoustique - Graduate School (IA-GS), CNRS, Le Mans Universit\'e, Av. Olivier Messiaen, 72085 Le Mans, France}
\author{Diego Bautista Avil\'es}
\email{diego.bautista@ug.uchile.cl}
\affiliation{Departamento de F\'{\i}sica, Facultad de Ciencias F\'{\i}sicas y Matem\'aticas, Universidad de Chile, Santiago 8370415, Chile}
\author{Luis E. F. Foa Torres}
\email{luis.foatorres@uchile.cl}
\affiliation{Departamento de F\'{\i}sica, Facultad de Ciencias F\'{\i}sicas y Matem\'aticas, Universidad de Chile, Santiago 8370415, Chile}
\author{Vassos Achilleos}
\email{Achilleos.Vassos@univ-lemans.fr}
\affiliation{Laboratoire d'Acoustique de l'Universit\'e du Mans (LAUM), UMR 6613, Institut d'Acoustique - Graduate School (IA-GS), CNRS, Le Mans Universit\'e, Av. Olivier Messiaen, 72085 Le Mans, France}

\date{\today}

 \begin{abstract}
Non-Hermitian systems driven along slow parametric loops undergo non-adiabatic transitions whose outcome depends sensitively on the driving speed, yet no explicit formula has been available for the critical timescale $T_{\mathrm{cr}}$ at which these transitions develop. Using a $2\times 2$ Hamiltonian with circular parameter trajectories, we derive $T_{\mathrm{cr}} = \mathcal{G}\,\ln(1/|\Delta|)$ in closed form for non-encircling loops, phase-shifted loops, offset loops, and loops encircling exceptional points, where $\mathcal{G}$ is a geometry-dependent growth factor and $\Delta$ is the instability seed. This formula sharply separates the regime where the system remains in the averagely dominant eigenstate ($T< T_{\mathrm{cr}}$) from the superadiabatic regime where the instantaneous dominant eigenstate takes over ($T> T_{\mathrm{cr}}$), resolving the apparent tension between the previous literature. We identify two competing seeds: a geometric Stokes multiplier and the finite-precision floor. When the geometric seed vanishes, precision alone governs the transition, yielding $T_{\mathrm{cr}} \propto m\ln\beta$, linear in the number $m$ of digits in base $\beta$ (bits for $\beta=2$). For loops encircling an exceptional point, the Stokes seed only sets an intermediate plateau; the final transition is again seeded by the precision floor, and $T_{\mathrm{cr}}$ remains linear in $m$. This provides a purely forward-evolution manifestation of precision-induced irreversibility (PIR)~\cite{PIR}, demonstrating that the fundamental limit identified through echo protocols also controls the outcome of slow non-Hermitian dynamics without requiring time reversal. For PT-symmetric energy spectra, $T_{\mathrm{cr}}$ additionally determines the onset of chirality: the dynamics is non-chiral for $T< T_{\mathrm{cr}}$ and chiral for $T> T_{\mathrm{cr}}$.
\end{abstract}

\maketitle

\section{Introduction} Non-Hermitian systems possess a remarkable feature absent from their Hermitian counterparts: degeneracies known as exceptional points (EPs), at which eigenvalues \emph{and} eigenvectors simultaneously coalesce \cite{Heiss1999,Heiss2000,Dembowski2001,Gao2015,Ding2016}. Because eigenstates are not orthogonal near an EP, a parameter loop threading its vicinity can exchange them, a topological effect observed in microwave cavities, exciton-polariton billiards, and acoustic resonators. More fundamentally, the non-orthogonality renders non-adiabatic transitions (NATs) \emph{unavoidable}: they persist even in the limit of infinitely slow driving \cite{FleischerMoiseyev2005,Uzdin2011,Gilary2013,IbanezMuga2014,Graefe2013}. 
 Two complementary frameworks have been advanced to explain this breakdown. Berry and Uzdin~\cite{Berry2011} traced it to the Stokes phenomenon, whereby subdominant exponentials switch on across Stokes lines regardless of driving speed. Milburn \emph{et al.}~\cite{Milburn2015}, working within dynamical bifurcation theory \cite{LebovitzSchaar1977,Neishtadt1987,Neishtadt1988,BerglundSchneider1999}, identified a stability-loss delay through which an infinitesimal perturbation is exponentially amplified during unstable phases. 

Following the terminology of Ref.~\cite{Nye2024}, the \emph{instantaneously dominant} eigenstate ($\mathcal{D}$) is defined as the state with the largest $\operatorname{Im} E(t)$ at each instant. In contrast, the \emph{averagely dominant} eigenstate ($\mathcal{D}_{\mathrm{av}}$) is the one that maximizes the accumulated gain, $\int_0^t \operatorname{Im} E(\tau)\, d\tau$. As discussed in Ref.~\cite{Nye2024}, the competition for dominance at any given time $t$ occurs between these two states, $\mathcal{D}$ and $\mathcal{D}_{\mathrm{av}}$.
In the limit of extremely slow evolution—referred to as the superadiabatic regime—the state that ultimately prevails is the instantaneously dominant one, $\mathcal{D}$, a point also made in ~\cite{Kumar2025} by Kumar \emph{et al.} However, for loops encircling an exceptional point (EP), the Stokes contribution alone is insufficient to fully drive the system onto the nonadiabatic branch; instead, it induces only a finite perturbation, which does not result in complete state conversion. Coming back to the extremely slow (superadiabatic) regime, a natural question is \emph{how slow is slow enough?} The practical importance of this question is backed by a decade of experiments demonstrating chiral state conversion, i.e.\ asymmetric final states depending on the traversal direction, in waveguides \cite{Doppler2016}, optomechanical systems \cite{Xu2016}, simple classical pendulums~\cite{Nenning25} and optical fiber loops \cite{Yoon2018}, including chiral transfer without encircling an EP~\cite{Nasari2022}, and by proposals linking chiral conversion to quantum measurement~\cite{foa_torres_non-hermitian_2025}. Analytical solutions for specific protocols~\cite{Hassan2017,Hassan2017a}, studies of starting-point and loop-geometry dependence~\cite{Zhang2018,Li2020,Feilhauer2020}, and general theories of universal~\cite{Nye2023,Nye2024} and slow~\cite{WangChong2018,Kumar2025} non-Hermitian evolution (see also  \cite{svegborn_framework_2026} extending this picture from non-Hermitian Hamiltonians to Liouvillian dynamics) have collectively established \emph{that} state conversion occurs and \emph{which} eigenstate ultimately wins.

Yet two questions central to both theory and experiment remain unanswered. First, no explicit formula exists for the critical timescale $T_{\mathrm{cr}}$ beyond which the non-adiabatic transition fully develops and the instantaneously dominant eigenstate takes over, despite $T_{\mathrm{cr}}$ being the quantity that determines the outcome of every slow non-Hermitian protocol. Second, the physical origin of the \emph{seed} $\Delta$ that initiates the exponential amplification is still debated: Berry and Uzdin attribute it to the Stokes phenomenon~\cite{Berry2011}, while Kumar \emph{et al.}\ attribute it to uncontrolled perturbations---modeled as external noise~\cite{Kumar2025}---but in neither case has a quantitative prediction for $T_{\mathrm{cr}}$ been obtained. These two questions are, in fact, linked by a deeper puzzle: perfect chiral conversion erases all memory of the initial state, yet the underlying dynamics consists of individually reversible integration steps~\cite{PIR}. 
Identifying the seed and the moment at which amplification renders the process irreversible are thus two facets of a single, deeper question: how does irreversibility emerge from nominally reversible evolution?

In this work we address both questions and bring light on the deeper puzzle. Using a $2\times 2$ non-Hermitian Hamiltonian with circular parameter trajectories, and building on the dynamical-bifurcation framework of Ref.~\cite{Milburn2015}, we derive explicit analytical expressions for $T_{\mathrm{cr}}$ across progressively general loop geometries: symmetric non-encircling loops, loops with shifted initial phase $\phi_0$ or offset center $g_0$, and loops encircling the EP. In every case the result takes the universal form \begin{equation}\label{Tcr_universal_intro} T_{\mathrm{cr}} = \mathcal{G}\,\ln\!\frac{1}{|\Delta|}\,, \end{equation} where $\mathcal{G}$ is a geometry-dependent growth factor that we compute in closed form. 
For parameter loops yielding PT-symmetric energy spectra, $T_{\mathrm{cr}}$ sharply separates non-chiral dynamics ($T< T_{\mathrm{cr}}$) from chiral dynamics ($T > T_{\mathrm{cr}}$), providing a quantitative criterion for the onset of chirality. We further identify the physical origin of $\Delta$. Two mechanisms can contribute: (i)~a geometric Stokes multiplier set by the asymptotic structure of the solution, and (ii)~the finite-precision floor of the computation or experiment. For symmetric non-encircling loops initialized in an eigenstate, the geometric seed vanishes and $\Delta$ reduces to $\beta^{-m}$, where $m$ is the number of mantissa digits and $\beta$ the base of the number representation ($\beta=2$ for binary arithmetic), so that $\beta^{-m}$ is the relative resolution of the implementation, yielding the parameter-free prediction $T_{\mathrm{cr}} = (2\pi/r^2)\,m\ln\beta$, linear in the number of precision bits~$m$. In an experimental realization, a noise floor $\varepsilon_{\mathrm{noise}}\gg\beta^{-m}$ plays exactly the same role, replacing the arithmetic floor with the same mechanism and the same geometry-dependent prefactor (Sec.~\ref{sec:seed}); the measured critical time then gives direct access to the effective resolution of the setup. The theory of precision-induced irreversibility (PIR)~\cite{PIR} established that amplification, non-normality, and finite precision combine to produce irreversibility in non-Hermitian dynamics, with a predictability horizon that scales linearly with the number of precision bits. Our results show that the same mechanism underlies the ``inevitable'' non-adiabatic transition: it is inevitable \emph{only because} every physical or computational platform operates at finite precision. The Hamiltonian provides the \emph{mechanism} for exponential amplification; precision provides the \emph{seed}. While PIR was originally identified through echo (time-reversal) protocols, the precision-dependent $T_{\mathrm{cr}}$ derived here demonstrates that the same fundamental limit governs purely forward evolution, manifesting as a concrete and previously unexplained observable: the critical timescale for state conversion.

The paper is organized as follows. Section~II introduces the model and notation. Section~III derives $T_{\mathrm{cr}}$ for non-encircling loops with progressively general geometries. Section~IV treats EP-encircling trajectories. Section~V analyzes the connection between $T_{\mathrm{cr}}$ and chiral dynamics. Section~\ref{sec:seed} investigates the physical origin of the seed~$\Delta$. We close with our conclusions in Section~VII. \section{Model and notation} We consider the time-dependent Schr\"odinger equation $\dot{\psi}=-iH\psi$ for a two-component state $\psi \in \mathbb{C}^2$ governed by the non-Hermitian Hamiltonian \begin{equation}\label{sy} H=\begin{pmatrix} iz&-1\\ -1&-iz \end{pmatrix} \end{equation} where $z=z(t)\in\mathbb{C}$ is a slowly varying control parameter. This is the canonical PT-symmetric dimer: the diagonal entries $\pm iz$ encode balanced gain and loss, while the off-diagonal entries couple the two modes. The system possesses two exceptional points at $z=\pm 1$, where eigenvalues \emph{and} eigenvectors simultaneously coalesce. The eigenvalues of $H$ are \begin{equation} E_{\pm}=\pm\sqrt{1-z^2}=\pm E \end{equation} and the corresponding right ($v_{\pm}$) and left ($u_{\pm}$) eigenvectors are organized as columns of $P$ and rows of $P^{-1}$, respectively: \begin{equation}\label{basis} \begin{split} &P=\begin{pmatrix} v_+&v_- \end{pmatrix}=\begin{pmatrix} \cos{\frac{\theta}{2}}&-\sin{\frac{\theta}{2}}\\ \sin{\frac{\theta}{2}}&\cos{\frac{\theta}{2}} \end{pmatrix}\\ &P^{-1}=\begin{pmatrix} u_+\\ u_- \end{pmatrix}=\begin{pmatrix} \cos{\frac{\theta}{2}}&\sin{\frac{\theta}{2}}\\ -\sin{\frac{\theta}{2}}&\cos{\frac{\theta}{2}} \end{pmatrix} \end{split} \end{equation} with $\theta$ defined by $\tan\theta=\frac{i}{z}$. Projecting onto this basis, $\phi=P^{-1}\psi=\begin{pmatrix} c_+&c_- \end{pmatrix}^T$, transforms the Schr\"odinger equation into \begin{equation}\label{system} \dot{\phi}=-i(D-iP^{-1}\dot{P})\phi,\quad D=\begin{pmatrix} E&0\\ 0&-E \end{pmatrix} \end{equation} which governs the mode amplitudes $c_{\pm}(t)$. The diagonal part $D$ drives exponential growth and decay, while the off-diagonal coupling $P^{-1}\dot{P}$, generated by the time dependence of the eigenbasis, is the source of non-adiabatic transitions between modes. We drive the system along circular parameter trajectories $z(t)=g_0-re^{-i(ft+\phi_0)}$ with period $T=2\pi/f$, radius~$r$, center offset $g_0\in\mathbb{R}$, and initial phase~$\phi_0$. In energy space, these loops cross the real axis---i.e.\ $\operatorname{Im} E(t)$ changes sign---at least once per period, exchanging the stability of the two modes and triggering the competition between adiabatic tracking and non-adiabatic transitions that we analyze below. In the following sections we derive $T_{\mathrm{cr}}$ for progressively more general parameter trajectories. In all cases, the result takes the universal form $T_{\mathrm{cr}} = \mathcal{G}\,\ln(1/|\Delta|)$, where $\mathcal{G}$ is a geometry-dependent growth factor and $\Delta$ is a seed amplitude that initiates the exponential amplification during unstable phases. Two mechanisms can contribute to $\Delta$: (i)~a geometric Stokes multiplier arising from the asymptotic structure of the solution, and (ii)~the finite-precision floor of the numerical or experimental implementation. We treat $\Delta$ as a fitting parameter through Secs.~III--V and systematically investigate its origin in Sec.~\ref{sec:seed}. Throughout, we characterize the evolution by the population ratio \begin{equation}\label{Rdef} R_{\pm}(t)=\frac{c_{\mp}(t)}{c_{\pm}(t)}, \end{equation} where the subscript labels the eigenstate in which the system is initialized: $|R_{\pm}|\ll1$ means the system still tracks $v_{\pm}$, while $|R_{\pm}|\gg1$ signals a transition to the complementary state. In Secs.~III and~IV the system is always initialized in the eigenstate whose adiabatic fixed point is unstable near the end of the period; which of $v_{\pm}$ this selects depends on the loop geometry and is stated in each case. Section~V, where chirality is analyzed, initializes the opposite state.
\section{Non-encircling loops} 
\begin{figure*}[t]
    \centering
\includegraphics[width=1\linewidth]{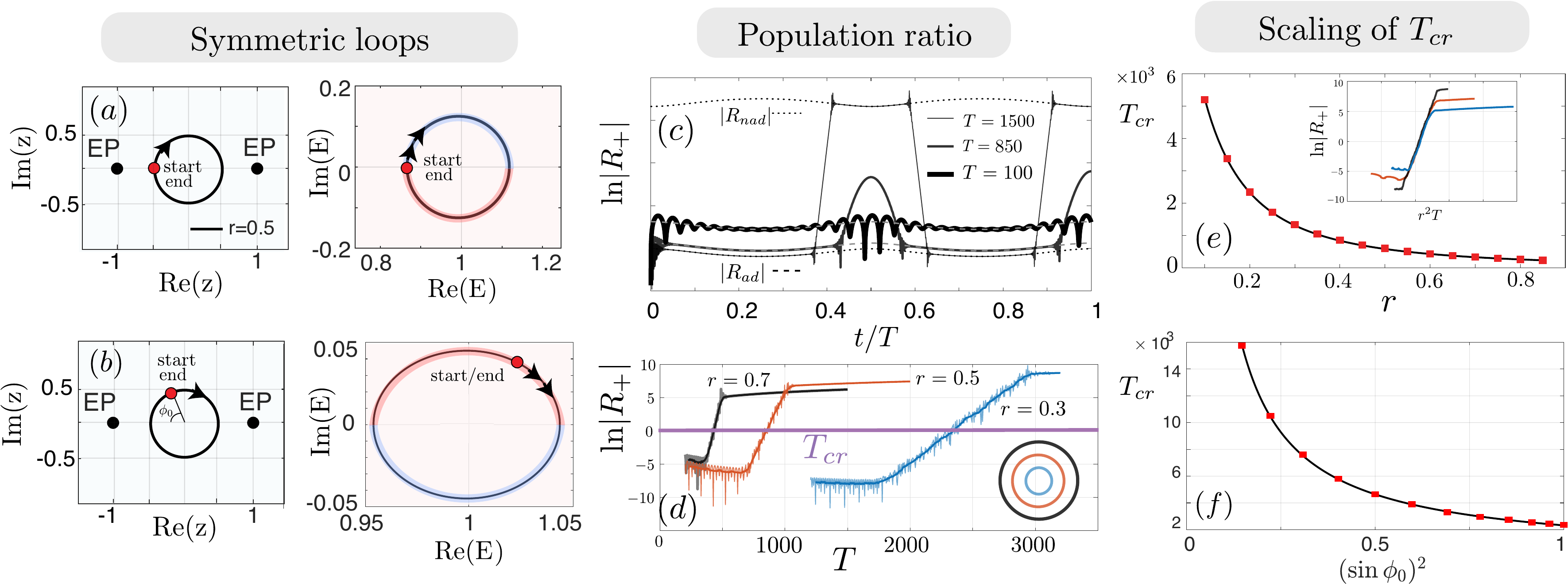}
 \caption{(a)~Parameter loop $z(t)=-r e^{-i f t}$, with exceptional points at $z=\pm1$. On the right the image in energy space, $E(t)=\sqrt{1-z^2(t)}$. The energy trajectory is traversed twice per period. Red markers denote the initial point ($t=0$), crossed again at $t=T/2$ and $t=T$. With blue we highlight the segment of the circle where $R_{ad}(t)$ is stable and with red when it is unstable. (b)~Parameter loop  $z(t)=-r e^{-i(f t+\phi_0)}$ for $t\in[0,T]$ and on the right its image under $E(t)=\sqrt{1-z^2(t)}$ where with blue/red we highlight the segments of the circle where $R_{ad}(t)$ is stable/unstable. (c) Time evolution of $\ln |R_+(t)|$ for the parameter loop of Fig.~1a and three periods: $T=100$, $850$, and $1500$. Dashed curves show the instantaneous adiabatic fixed points $|R_{\mathrm{ad}}(t)|$; the dotted curve shows $|R_{\mathrm{nad}}(t)|$ for $T=1500$. The Schr\"odinger equation is solved using a fourth-order Runge--Kutta scheme with time step $dt = 10^{-2}$; the coefficients $c_{\pm}(t)$ are obtained by projecting $\psi(t)$ onto the left eigenvectors $u_{\pm}$ of Eq.~\eqref{basis}. (d) $\ln |R_+(T)|$ as a function of $T$ for three loop radii $r = 0.3$, $0.5$, and $0.7$, with $z(t) = -r e^{-i f t}$. The horizontal purple line marks $|R_+| = 1$, defining $T_{\mathrm{cr}}$. (e)~Fit of $\text{const.}/r^2$ to the numerically extracted $T_{\mathrm{cr}}$ vs $r$ (red markers). Inset: rescaling the horizontal axis to $r^{2}T$ collapses the transition regions. (f)~Fitted curve to the numerically extracted $T_{\mathrm{cr}}$ vs $\phi_0$ for $r=0.3$. The functional form is $\text{const.}/(r\sin\phi_0)^2$ [Eq.~\eqref{res2}].} 
\label{fig:Figure1}
\end{figure*}
We study Eq.~\eqref{system} for the non-encircling loops $z(t) =g_0-re^{-i(ft+\phi_0)}$, see Fig.~\ref{fig:Figure1}(a,b) and Fig.~\ref{fig:Figure2}(a,b), closely following the analysis of Ref.~\cite{Milburn2015} (see also Appendix~\ref{app:db}). 
\subsection{Symmetric loop without phase shift}
We first consider $g_0=0$ and $\phi_0=0$. The system is initialized in $\psi(0)=v_{+}(0)$, so the relevant quantity is the population ratio $R_+(t)=c_-(t)/c_+(t)$, which satisfies the Riccati equation (mapping the linear flow on $\mathbb{C}^2$ to a nonlinear flow on $\mathbb{CP}^1$): \begin{equation}\label{wxx} \dot{R}_{+}=2iE(t)R_{+}+ h(t)(1+R_{+}^2),\quad R_+(0)=0 \end{equation} with $h(t)=rfe^{-ift}/(2(1-r^2e^{-2ift}))$. As long as $|R_{+}(t)|\ll 1$ the system remains in the initialized state $v_+(t)$, while $|R_+(t)|\gg 1$ signals a transition to $v_-(t)$. Equation~\eqref{wxx} admits two instantaneous fixed points \begin{equation}\label{fp} R_{ad}(t)\approx\frac{ih(t)}{2E(t)},\quad R_{nad}(t)\approx \frac{2E(t)}{ih(t)} \end{equation} In the adiabatic regime, $|R_{ad}(t)|\ll 1$ and $|R_{nad}|\gg 1$, so proximity to each fixed point indicates adiabatic and non-adiabatic behavior, respectively. The stability of the fixed points depends on the sign of $\Im(E(t))$: $R_{ad}(t)$ is stable for $\Im(E(t))>0$ and $R_{nad}(t)$ for $\Im(E(t))<0$. At the times $t_*$ where $\Im(E(t_*))=0$, the stability is exchanged. Fig.~\ref{fig:Figure1}(a) illustrates this: blue segments mark intervals where $R_{\mathrm{ad}}(t)$ is stable, $t \in (0,T/4) \cup (T/2,3T/4)$, and red segments mark unstable intervals, $t \in (T/4,T/2) \cup (3T/4,T)$. Fig.~\ref{fig:Figure1}(c) shows numerical solutions of $\ln|R_+(t)|$ for three values of the period $T$, all satisfying the adiabaticity condition $|h(t)/2E(t)|\ll 1$.

For $T=100$, the solution $R_+(t)$ tracks $R_{ad}(t)$ throughout, indicating adiabatic evolution. For $T=850$, the system begins to populate $v_-(t)$ whenever $R_{ad}(t)$ becomes unstable, reaching maximum population at $t=T/2$ and $t=T$, before reverting to $R_{ad}(t)$ as stability is restored. For $T=1500$, the solution reaches $R_{nad}(t)$, completing a non-adiabatic transition to $v_-(t)$ where it remains for a significant duration. The deviation from the adiabatic fixed point, and the time the solution spends near either fixed point, depend sensitively on $T$ at fixed~$r$. This $T$-dependence is summarized in Fig.~\ref{fig:Figure1}(d), where $\ln|R_+(T)|$ is plotted as a function of $T$ for several radii~$r$. Each curve exhibits two plateaus corresponding to the adiabatic and nonadiabatic fixed points. We define the critical timescale $T_{cr}$ as the period at which the two states are equally populated at the end of the evolution, i.e.\ $|R_+(T_{cr})|=1$. This divides the parameter space into three regions: for $T<T_{cr}$ the system ends in the initialized state; near $T\approx T_{cr}$ the state $v_-(t)$ begins to populate; and for $T>T_{cr}$ the state $v_-$ fully dominates. The transition between the two plateaus is approximately linear in~$T$, and shifts to larger values of~$T$ as $r\to 1$.
To derive $T_{cr}$ analytically, we use the linearized solution of Eq.~\eqref{wxx}, derived in Appendix~\ref{app:db} [Eq.~\eqref{wwx}]: \begin{equation}\label{si} R_+(t)\approx\mathcal{R}^{(N)}_{ad}(t)-\mathcal{R}^{(N)}_{ad}(0)e^{W(t)}+\Delta(t) e^{\Psi(t)} \end{equation} where \begin{equation}\label{int} W(t)=2i\int_0^tE(\tau)d\tau,\quad \Psi(t)=2i\int_{t_*}^tE(\tau)d\tau \end{equation} For $r\ll 1$ the energy trajectory is approximately circular (Fig.~\ref{fig:Figure1}(a)), giving \begin{equation}\label{en} E(t)\approx 1-(r^2/2)e^{-2ift} \end{equation} The first two terms in Eq.~\eqref{si} accurately describe the evolution for all $T<T_{cr}$, yielding purely adiabatic dynamics (see App.~\ref{app:accuracy}). The last term is therefore negligible below $T_{cr}$. For $T\gtrsim T_{cr}$, the last term drives the deviation from $\mathcal{R}_{\mathrm{ad}}(t)$. At $t_*=3T/4$, the closest root of $\Im E(t)=0$ to $T$, the adiabatic fixed point becomes unstable, and the remainder can be approximated as \begin{equation}\label{okk} \Delta(t) e^{\Psi(t)}\approx\Delta\Theta(t-t_*)e^{\Psi(t)},\quad t_*=\frac{3T}{4} \end{equation} where $\Delta$ is a time-independent function of the system parameters and \begin{equation} \Psi(t)=2i(t-t_*)+\frac{r^2}{2f}(e^{-2ift}-e^{-2ift_*}) \end{equation}
The seed's origin is the remainder of the linearized solution derived in App.~\ref{app:db}: the first two terms of Eq.~\eqref{si} are fixed by the endpoints and cannot by themselves produce a transition, so any departure from adiabatic tracking must come from the last term, whose amplitude at the moment stability is exchanged is $\Delta$. Whether it is set by the geometry of the loop or by the resolution floor of the implementation is the question taken up in Sec.~\ref{sec:seed}. Using Eq.~\eqref{dt} with the assumption $\Delta\sim\text{const.}$, we obtain the central result of this section: \begin{equation}\label{res1} T_{cr}\approx\frac{2\pi}{r^2}\ln\frac{1}{|\Delta|} \end{equation} The physical origin of $\Delta$, whether geometric or precision-related, is analyzed in Sec.~\ref{sec:seed}. Fig.~\ref{fig:Figure1}(e) confirms this prediction: the fitted curve $T_{cr}(r)$ agrees with Eq.~\eqref{res1}, and rescaling the horizontal axis to $Tr^2$ (inset of Fig.~\ref{fig:Figure1}(e)) collapses the transition curves for different values of~$r$. In summary, for times near the end of the evolution ($t\in(3T/4,T)$), $\Im E(t)<0$ and $R_{ad}(t)$ is unstable. Whether the system has enough time to complete the transition to $R_{nad}(t)$ depends on the delay time (App.~\ref{app:db}): the parameters $(r,T)$ must allow the exponential amplification to overcome the seed $\Delta$ before the period ends. The result is \begin{equation} \begin{split} v_+(0)\to v_+(T),\quad T<T_{cr}\\ v_+(0)\to v_-(T),\quad T> T_{cr} \end{split} 
\end{equation}
\subsection{Phase-shifted symmetric loop}
Without loss of generality, we restrict to $\phi_0\in(0,\pi/2]$. The relevant solution is $R_-(t)$, corresponding to the initial condition $\psi(0)=v_-(0)$; Fig.~\ref{fig:Figure1}(b) highlights in blue and red the segments of the energy-space trajectory where $R_{\mathrm{ad}}(t)$ is stable and unstable, respectively. Following the same reasoning as in Sec.~III\,A, for $T<T_{\mathrm{cr}}$ the first two terms of the solution $R_-(t)$ accurately reproduce the numerical results, see App.~\ref{app:accuracy}. Unlike the $\phi_0=0$ case, the trajectory geometry causes these two terms to deviate from $\mathcal{R}_{\mathrm{ad}}(t)$ at intermediate times (Fig.~\ref{fig:Figure6}(b), shown in the appendices). However, near $t\sim T$, where $\Re[W(t\to T)]\to 0$, they cannot describe the NAT to $v_+(T)$ observed for $T\gtrsim T_{\mathrm{cr}}$. The term responsible for the transition is therefore $\Delta(t)e^{\Psi(t)}\approx\Delta\Theta(t-t_*)e^{\Psi(t)}$, with $t_*=T(1-\phi_0/2\pi)$ the closest root of $\Im{E(t)}=0$ to $T$, and \begin{equation} \Psi(t)=-2i(t-t_*)-\frac{r^2}{2f}(e^{-2i(ft+\phi_0)}-e^{-2i(ft_*+\phi_0)}) \end{equation} Using the definition of $T_{cr}$ we obtain the expression \begin{equation}\label{res2} T_{cr}\approx\frac{2\pi}{(r\sin\phi_0)^2}\ln\left|\frac{1}{\Delta}\right|, \end{equation} where $r\sin\phi_0$ is the vertical distance of the initial point on the circular trajectory from the real axis. In deriving Eq.~\eqref{res2}, we again assume $\Delta\approx\text{const}$. Figure~\ref{fig:Figure1}(f) confirms this scaling: the fitted curve agrees well with Eq.~\eqref{res2}. Note that $T_{\mathrm{cr}}\to\infty$ as $\phi_0\to 0$; this does not contradict Sec.~III\,A, since Eq.~\eqref{res2} assumes initialization in $v_-(0)$. 
\subsection{Off-centered loops}
\begin{figure*}
    \centering
    \includegraphics[width=1\linewidth]{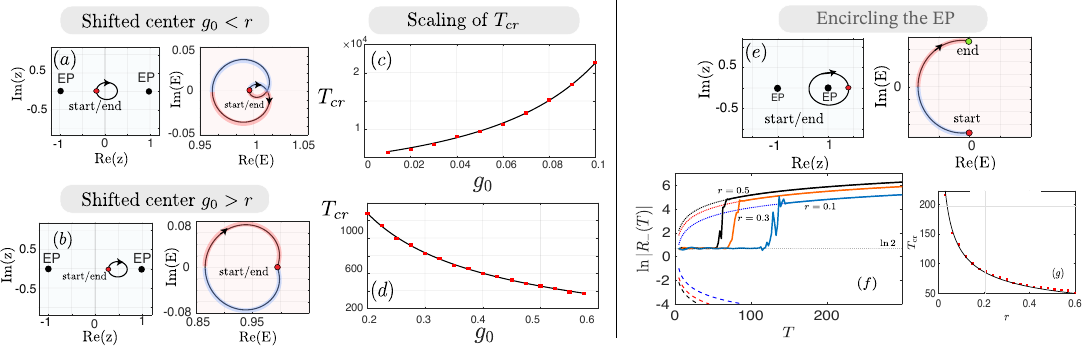}
    \caption{(a)~Parametric curve $z(t)=g_0-re^{-ift}$ for $g_0<r$. (b)~Same for $g_0\geq r$; red markers indicate the start/end point. (c)~Fitted curve to the numerically extracted $T_{\mathrm{cr}}$ vs.\ $g_0$ for $g_0<r$ and $r=0.2$; the functional form is $\text{const.}\,(1-g_0^2)^{1/2}/(r-g_0)^2$ [Eq.~\eqref{res3}]. (d)~Same for $g_0\geq r$; the functional form is $\text{const.}\,(1-g_0^2)^{1/2}/(g_0 r)$ [Eq.~\eqref{res4}]. (e)~Parametric curve $z(t)=1+re^{-ift}$ (EP-encircling) and its image in energy space on its right. (f)~$\ln|R_-(T)|$ vs.\ $T$ for $z(t)=1+re^{-ift}$ and three loop radii: $r=0.1$ (blue), $0.3$ (orange), and $0.5$ (black). Dashed and dotted curves show $|R_{\mathrm{ad}}(T)|$ and $|R_{\mathrm{nad}}(T)|$, respectively, for each~$r$. (g)~Fitted curve to the numerically extracted $T_{\mathrm{cr}}$ vs $r$. The functional form is $\text{const.}/(r)^{1/2}$.}
    \label{fig:Figure2}
\end{figure*}

We now consider the circular trajectories $z(t)=g_0-re^{-ift}$, see Figs.~\ref{fig:Figure2}(a,b). In this case we approximate the energy as \begin{equation}\label{gene} \begin{split} E(t)&\approx(1-g_0^2)^{1/2}+\frac{g_0r}{(1-g_0^2)^{1/2}}e^{-ift}\\ &-\frac{r^2}{2(1-g_0^2)^{1/2}}e^{-2ift} \end{split} \end{equation} This approximation breaks down for $r+g_0\geq 1$, i.e.\ when the trajectory encircles the EP. As in the previous section, the system is initialized in the eigenstate for which $R_{\mathrm{ad}}(t)$ is unstable near the end of the evolution.

We consider separately the cases $g_0<r$ and $g_0\geq r$, for which the relevant solutions are $R_+(t)$ and $R_-(t)$, respectively. Figs.~\ref{fig:Figure2}(a,b) highlight in blue/red the stable/unstable segments of the energy-space trajectories. For $g_0<r$, the same analysis gives \begin{equation} \Psi(t)=a(t-t_*)-b(e^{-ift}-e^{-ift_*})+c(e^{-2ift}-e^{-2ift_*}) \end{equation} where \begin{equation} \begin{split} &a=2i(1-g_0^2)^{1/2}\\ &b=\frac{2g_0r}{f(1-g^2_0)^{1/2}}\\ &c=\frac{r^2}{2f(1-g_0^2)^{1/2}} \end{split} \end{equation} and $t_*$ is the root of $\cos(ft)=g_0/r$ closest to $T$. This yields \begin{equation}\label{res3} T_{cr}\approx\frac{2\pi(1-g_0^2)^{\frac{1}{2}}}{(r-g_0)^2}\ln\frac{1}{|\Delta|},\quad g_0<r \end{equation} 
For $g_0\geq r$, we find \begin{equation}
\Psi(t)=-a(t-t_*)+b(e^{-ift}-e^{-ift_*})-c(e^{-2ift}-e^{-2ift_*}) \end{equation} with $t_*=T/2$. This gives \begin{equation}\label{res4} T_{cr}\approx\frac{\pi(1-g_0^2)^{1/2}}{2g_0r}\ln\frac{1}{|\Delta|},\quad g_0\geq r \end{equation} 
\section{Encircling the EP}\label{Encircling} We now set $g_0 = 1$, so that the circular trajectory encloses the EP at $+1$, see Fig.~\ref{fig:Figure2}(e). To facilitate comparison with Ref.~\cite{Berry2011}, we restrict to the case $\phi_0=\pi$. The dependence of $T_{\mathrm{cr}}$ on the number of precision digits for this geometry is shown in Fig.~\ref{fig:Tcr_vs_bits} and analyzed in Sec.~\ref{sec:seed}. The system is initialized in $v_-(0)$, so the relevant quantity is $R_-(t)$. Fig.~\ref{fig:Figure2}(f) shows $|R_-(T)|$ for three values of~$r$. A qualitative difference from the non-encircling case appears in the first plateau: $|R_-(T)|$ does not relax to $|R_{\mathrm{ad}}(T)|$ but saturates at $|R_-(T)| \approx 2$, a value independent of $r$ and~$T$. As discussed in Appendix~\ref{app:EPenc} and analyzed in Refs.~\cite{Berry2011,Milburn2015}, this behavior originates from the Stokes phenomenon, which becomes active at $t = T/2$ upon encircling the EP. For larger~$T$, the solution undergoes a sharp transition to $|R_{\mathrm{nad}}(T)|$: for $t > T/2$ the non-adiabatic branch is dynamically stable, and the system ends in the complementary eigenstate. However, when the intrinsic basis flip from EP encirclement is accounted for, the final physical state coincides with the initial one. To determine $T_{\mathrm{cr}}$, we approximate the energy as $E(t)=-i\sqrt{2r}e^{-ift/2}$ and use Eq.~\eqref{dt} with \begin{equation} \Psi(t)=-\frac{4i\sqrt{2r}}{f}(e^{-\frac{i}{2}ft}-e^{-\frac{i}{2}ft_*}),\quad t_*=\frac{T}{2} \end{equation} This yields \begin{equation}\label{Tcr_EP} T_{cr}\approx \frac{\pi}{2\sqrt{2r}}\ln{\frac{1}{|\Delta|}} \end{equation} Figure~\ref{fig:Tcr_vs_bits} was computed with Berry's local model \[ \dot{x}=\begin{pmatrix}0&1\\ \rho e^{-ift}&0\end{pmatrix}x \] \cite{Berry2011}, the near-EP limit of Eq.~\eqref{sy} with $\rho=2r$; there $r$ denotes $\rho$, and Eq.~\eqref{Tcr_EP} becomes $T_{\mathrm{cr}}\approx[\pi/(2\sqrt{\rho})]\ln(1/|\Delta|)$, which for $|\Delta|=2^{-m}$ gives the slope quoted in its caption. Fig.~\ref{fig:Figure2}(g) confirms the characteristic $r^{-1/2}$ dependence. The final-state assignment is therefore \begin{equation}\label{stokesresult}
\begin{split} &v_-(0)\to c_+v_+(T)+c_-v_-(T),\quad T<T_{cr}\\ &v_-(0)\to v_+(T),\quad T> T_{cr} \end{split} \end{equation} where $|c_+/c_-|=2$.

\section{Chiral conversion and the role of $T_{\mathrm{cr}}$} 
\begin{figure}[ht] \includegraphics[width=1\linewidth]{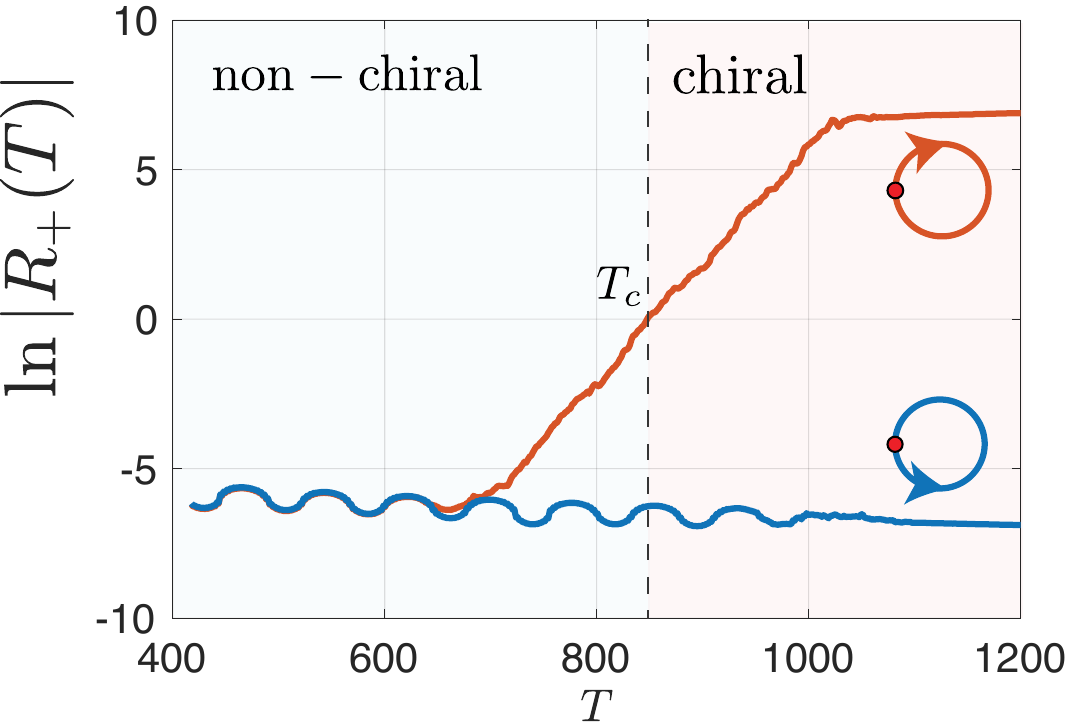} \caption{$\ln|R_+(T)|$ for clockwise (red) and anticlockwise (blue) dynamics, confirming Eq.~\eqref{chiral}.}
\label{fig:Figure3}
\end{figure}

\begin{figure}[ht] \includegraphics[width=1\linewidth]{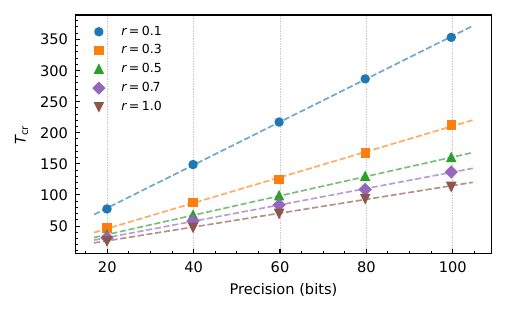} \caption{Critical timescale $T_{\mathrm{cr}}$ as a function of precision bits $m$ for EP-encircling loops. Data points are obtained from simulations using arbitrary-precision arithmetic for different radii. The nearly linear scaling with slope $\ln(2)\,\pi/(2\sqrt{r})$ is consistent with precision-limited behavior. } \label{fig:Tcr_vs_bits} \end{figure}

Several studies have examined whether the final state after one period depends or not on the encirclement direction, leading to chiral or non-chiral behavior respectively. We begin with the parametric loop $z(t)=-re^{-ift}$. Initializing in $\psi(0)=v_-(0)$, we study $R_-(t)=c_+(t)/c_-(t)$. Note that here, unlike in Secs.~III and~IV, the system is deliberately initialized in the state whose adiabatic fixed point is \emph{stable} near the end of the period. Numerical simulations show that, for arbitrary $T$, the system returns to $v_-$ after one period, consistent with $R_{\mathrm{ad}}(t \approx T)$ being stable. Using the ODE~\eqref{wx} and the PT-symmetry of the energy, $E^*(-ft)=E(ft)$, one finds that $R_+^*(t)$ for $f<0$ and $R_-(t)$ for $f>0$ satisfy the same equation with the same initial conditions. Consequently, \begin{equation}\label{symm} |R_+(-ft)|=|R_-(ft)| \end{equation} This implies that the evolution from $v_+(0)$ under clockwise encirclement equals that from $v_-(0)$ under anticlockwise encirclement; hence an anticlockwise evolution initialized in $v_+(0)$ necessarily returns to the same state. Combining these results (confirmed numerically in Fig.~\ref{fig:Figure3}): \begin{equation}\label{chiral} \begin{split} &v_+(0)\overset{\circlearrowright}{\longrightarrow}v_+(T),\quad v_+(0)\overset{\circlearrowleft}{\longrightarrow}v_+(T),\quad T< T_{cr}\\ &v_+(0)\overset{\circlearrowright}{\longrightarrow}v_-(T),\quad v_+(0)\overset{\circlearrowleft}{\longrightarrow}v_+(T),\quad T> T_{cr} \end{split} \end{equation} leading to the statement \begin{equation}\label{xa} \begin{split} &T< T_{cr}\,,\quad \text{The dynamics is non-chiral}\\ &T> T_{cr}\,,\quad\text{The dynamics is chiral} \end{split} \end{equation} The same result holds for $g_0\neq 0$, $\phi_0=0$, since the energy eigenvalues remain PT-symmetric. When $\phi_0\neq 0$, however, this symmetry is broken and chirality must be assessed case by case. For the EP-encircling loop of Sec.~\ref{Encircling}, $E^*(-ft)=-E(ft)$ which implies $R^{*}_{\pm}(-ft)=R_{\pm}(ft)$, so the dynamics is always non-chiral, in agreement with Ref.~\cite{Zhang2018}. That work also reported an experimental observation of Eq.~\eqref{xa} in a coupled ferromagnetic waveguide system: sufficiently long waveguides always produce chiral dynamics, but the waveguide used was not long enough to trigger the NAT required for chirality. Chiral state transfer without encircling an EP has also been observed experimentally, in a gain--detuning loop that becomes chiral only when traversed slowly enough~\cite{Nasari2022}; for the PT-symmetric loops considered here, Eq.~\eqref{xa} places that onset at $T_{\mathrm{cr}}$. 

An independent analysis appearing after this work, by Wang~\cite{wang_chiral_2026}, reaches the same conclusion for the time-periodic loops of Ref.~\cite{Zhang2018} and provides a parallel confirmation of PIR~\cite{PIR}.
  
  \section{The origin of the instability seed}\label{sec:seed} Throughout Secs.~III--V, the critical timescale takes the universal form $T_{\mathrm{cr}} = \mathcal{G}\,\ln(1/|\Delta|)$, where $\mathcal{G}$ is a geometry-dependent growth factor and $\Delta$ was treated as a fitting parameter. What determines $\Delta$ physically? As anticipated in Sec.~II, two mechanisms can contribute: (i)~a geometric Stokes multiplier $\Delta_{\mathrm{geo}}$, set by the asymptotic structure of the solution, and (ii)~the finite-precision floor $\Delta_{\mathrm{fp}} \sim \beta^{-m}$ of the computation or experiment. (Truncation errors from time-stepping constitute a third, purely numerical source; we verify that they are subdominant by confirming convergence with respect to the time step.) We now examine the two physical mechanisms in turn.
  
  \subsection{Symmetric non-encircling loops: precision as the sole seed} 
\begin{figure}[h] \includegraphics[width=1\linewidth]{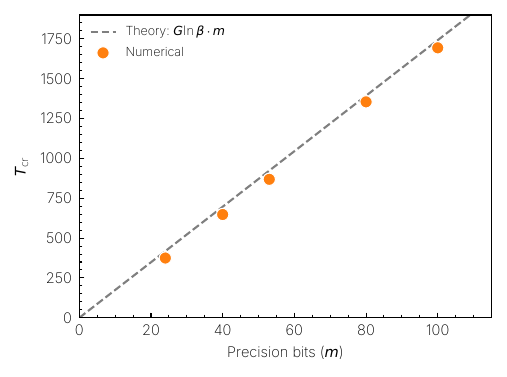} \caption{Critical timescale $T_{\mathrm{cr}}$ as a function of precision bits $m$ for symmetric non-encircling loops. Data points (circles) are obtained from simulations using arbitrary-precision arithmetic. The dashed line shows the theoretical prediction $T_{\mathrm{cr}} = \mathcal{G}\, m \ln\beta$ from Eq.~\eqref{Tcr_precision} with $\mathcal{G} = 2\pi/r^2$, yielding excellent agreement and confirming the linear scaling.} \label{fig:Tcr_vs_m} \end{figure}

For the symmetric loop ($g_0 = 0$, $\phi_0 = 0$), the geometric Stokes seed vanishes: $\Delta_{\mathrm{geo}} = 0$. Yet non-adiabatic transitions still occur. The seed must therefore originate from the precision floor $\beta^{-m}$. This identification yields a parameter-free prediction. Substituting $|\Delta| = \beta^{-m}$ into Eq.~\eqref{res1} gives \begin{equation}\label{Tcr_precision} T_{\mathrm{cr}} = \frac{2\pi}{r^2} \cdot m \ln\beta \end{equation} which is linear in the number of precision bits $m$. Figure~\ref{fig:Tcr_vs_m} confirms this scaling: the measured slope agrees with the theoretical prediction $\mathcal{G}\ln\beta$ with $\mathcal{G} = 2\pi/r^2$, confirming the linear dependence on precision bits. This result establishes that the ``inevitable'' non-adiabatic transition is inevitable \emph{only given finite precision}---with $m \to \infty$, we would have $T_{\mathrm{cr}} \to \infty$ and no transition at all. Using information-theoretic arguments, Del Santo and Gisin have argued that attributing infinitely many predetermined digits to a physical quantity is itself unphysical~\cite{DelSanto2019,Gisin2020}; for a recent argument in the context of chaotic dynamics---that the perturbation from any device following the motion renders determinism ``irrelevant to physics''---see Ref.~\cite{Berry2025}. Appendix~\ref{app:coeffs} examines the same transition at the level of the coefficient magnitudes, showing that it is carried by the population ratio while the coefficients themselves never become anomalously small. 
  
  \subsection{EP-encircling loops: Stokes vs precision} 
For loops encircling an EP, the Stokes phenomenon generates a finite geometric seed $\Delta_{\mathrm{geo}} \neq 0$. In the case studied in Sec.~\ref{Encircling} the evolution proceeds in two stages, clearly visible in Fig.~\ref{fig:Figure2}(f). For $T < T_{\mathrm{cr}}$, the population ratio settles at a plateau $|R_-(T)| \approx 2$, which is a direct signature of the Stokes contribution (App.~\ref{app:EPenc} derives the Stokes multiplier). In this regime, the dynamics is reversible and the state which prevails is given by Eq.~\eqref{stokesresult}. For $T > T_{\mathrm{cr}}$, a subsequent jump from the plateau at $|R_-(T)|\approx 2$ to $R_{\mathrm{nad}}$ occurs, driven by exponential amplification of the seed. This second stage marks the onset of irreversibility: the system transitions to the instantaneously dominant eigenstate and cannot spontaneously return to the original branch. As we saw the critical time scale is  $T_{\mathrm{cr}} = [\pi/(2\sqrt{2r})]\ln(1/|\Delta|)$ with the characteristic $r^{-1/2}$ scaling, in contrast to the $r^{-2}$ dependence of non-encircling loops, reflecting the topological distinction (monodromy: $\sqrt{z} \to -\sqrt{z}$ after one period).  
  
   \subsection{Hermitian starting point: absence of Stokes and $\Delta_{fp}$} The Hermitian starting point, \( z(t) = 1 - r e^{-i f t} \), represents a distinct limiting case. The corresponding trajectory in energy space begins on the positive real axis and terminates on the negative real axis. As a result, no Stokes line is crossed during the encircling, and the Stokes mechanism is therefore absent, i.e., \( \Delta_{\mathrm{geo}} = 0 \).

Moreover, there exists no time \( t_* \), since \( \operatorname{Im} E(t) > 0 \) for all \( t \in (0, T) \). Consequently, \( \Delta_{fp} = 0 \), and any exponential growth or decay is governed solely by \( W(t) \). Using Ref.~\cite{Nye2024} terminology, there is no competition between the averagely dominant state and the instantaneous one. Such competition can arise only if at least one \( t_* \) exists. As a result, the precision floor is irrelevant, and the dynamics remain fully reversible.

  \subsection{Distinguishing Stokes-limited from precision-limited behavior} The critical test for distinguishing the two different types of behavior is whether $T_{\mathrm{cr}}$ depends on precision. Figure~\ref{fig:Tcr_vs_bits} shows $T_{\mathrm{cr}}$ as a function of precision bits $m$ for EP-encircling loops at several radii. A linear dependence $T_{\mathrm{cr}} \propto m$ indicates that precision dominates (PIR-limited), while a plateau would indicate that the geometric Stokes seed dominates (Stokes-limited). \subsection{Unified picture} The effective seed is determined by whichever mechanism provides the largest initial perturbation: \begin{equation}\label{Delta_eff} \Delta_{\mathrm{eff}} = \max\!\left(\Delta_{\mathrm{geo}},\; \Delta_{\mathrm{fp}}\right) \end{equation} yielding the general formula \begin{equation}\label{Tcr_general} T_{\mathrm{cr}} = \mathcal{G}\,\ln\!\left(\frac{1}{|\Delta_{\mathrm{eff}}|}\right) \end{equation} from which all special cases derived in Secs.~III--V follow as evaluations of $\mathcal{G}$. The ``inevitable'' non-adiabatic transition is inevitable given a finite seed, and any real computation or experiment always has one. The precision floor $\Delta_{\mathrm{fp}} \sim \beta^{-m}$ corresponds to the resolution floor $\varepsilon$ of Ref.~\cite{PIR}; when experimental noise $\varepsilon_{\mathrm{noise}} \gg \beta^{-m}$ dominates, it simply replaces $\Delta_{\mathrm{fp}}$ in Eq.~\eqref{Delta_eff}, yielding an earlier transition with the same mechanism.  
  
  \subsection{The transition as an irreversibility signature} The jump from $R_{\mathrm{ad}}(t)$ to $R_{\mathrm{nad}}(t)$ is itself an irreversible event: once precision-induced errors trigger the transition, the system cannot spontaneously return to its original branch. This provides an alternative probe of precision-induced irreversibility (PIR) \cite{PIR}---rather than measuring fidelity loss after a time-reversal echo, one observes the precision-dependent threshold $T_{\mathrm{cr}}$ at which forward evolution itself becomes irreversibly altered. The state transition constitutes direct evidence of irreversibility without requiring an echo protocol.   \begin{table*}[t]
\centering
\begin{tabular}{c c c}
\hline\hline
Geometry & Condition & Power-law scaling of $T_{\mathrm{cr}}$ \\
\hline
Symmetric loop without phase shift & $g_0=0,\;\phi_0=0$ & $T_{\mathrm{cr}} \propto r^{-2}$ \\
Phase-shifted symmetric loop & $g_0=0,\;\phi_0\neq0$ & $T_{\mathrm{cr}} \propto (r\sin\phi_0)^{-2}$ \\
Off-centered loop & $g_0<r$ & $T_{\mathrm{cr}} \propto (1-g^2_0)^{1/2}(r-g_0)^{-2}$ \\
Off-centered loop& $g_0\ge r$ & $T_{\mathrm{cr}} \propto (1-g^2_0)^{1/2}(g_0 r)^{-1}$ \\
EP-encircling loop & $g_0=1$ & $T_{\mathrm{cr}} \propto r^{-1/2}$ \\
\hline\hline
\end{tabular}
\caption{Universal power-law scaling of the critical time $T_{\mathrm{cr}}$ for different circular parameter trajectories $z(t)=g_0-re^{-i(ft+\phi_0)}$. The full expressions take the general form $T_{\mathrm{cr}}=\mathcal{G}\ln(1/|\Delta|)$, where the geometry-dependent factor $\mathcal{G}$ produces the power laws shown above.}
\label{tab:Tcr_scaling_power}
\end{table*}

\section{Final Remarks and Discussion}

The introduction posed two open questions: what is the critical timescale $T_{\mathrm{cr}}$ beyond which non-adiabatic transitions fully develop, and what seeds the exponential amplification that drives them? Our analysis answers both. In every loop geometry examined, the critical timescale takes the universal form $T_{\mathrm{cr}} = \mathcal{G}\,\ln(1/|\Delta|)$, where $\mathcal{G}$ is a geometry-dependent growth factor computed in closed form (Table~\ref{tab:Tcr_scaling_power}) and $\Delta$ is the instability seed. This formula resolves the apparent tension between the Berry--Uzdin and Nye--Kumar predictions. Using the terminology of Ref.~\cite{Nye2024}, $T_{\mathrm{cr}}$ marks the boundary between the $\mathcal{D}_{av}$ regime ($T< T_{\mathrm{cr}}$), where the averagely dominant eigenstate prevails, and the $\mathcal{D}$ regime [$T> T_{\mathrm{cr}}$, termed superadiabatic, where the instantaneous dominant eigenstate takes over; see Figs.~\ref{fig:Figure1}(e,f) and \ref{fig:Figure2}(c,d,g)]. Both pictures are correct in their respective regimes. The physical mechanism is transparent in the Riccati formulation: along a single period, the stability of the adiabatic fixed point $R_{\mathrm{ad}}(t)$ is exchanged at multiple times $t_*$, and whether the solution can escape the now-unstable branch before stability is restored depends on whether the period $T$ exceeds $T_{\mathrm{cr}}$. For PT-symmetric energy spectra ($\phi_0 = 0$), $T_{\mathrm{cr}}$ also determines the onset of chirality [Eq.~\eqref{xa}]: the dynamics is non-chiral for $T< T_{\mathrm{cr}}$ and chiral for $T> T_{\mathrm{cr}}$.

The formula also reveals the physical origin of the seed $\Delta$. Two mechanisms compete: a geometric Stokes multiplier $\Delta_{\mathrm{geo}}$, set by the asymptotic structure of the solution, and the finite-precision floor $\Delta_{\mathrm{fp}} \sim \beta^{-m}$ of the computation or experiment. When the geometric seed vanishes, as for symmetric non-encircling loops initialized in an eigenstate, precision alone governs the transition, yielding the parameter-free prediction $T_{\mathrm{cr}} = \mathcal{G}\,m\ln\beta$. Unlike the noise-driven mechanism of Ref.~\cite{Kumar2025}, which requires external perturbations with tunable parameters, the precision floor is universal and fixed once the number of bits $m$ is specified.

These findings illuminate a deeper question raised in the introduction: how does irreversibility emerge from individually reversible integration steps? The general answer was given by PIR theory~\cite{PIR}, which showed that amplification, non-normality, and finite precision form a sufficient trinity for irreversibility, with a predictability horizon scaling linearly with precision bits. Our analysis reveals the specific mechanism through which this general principle operates in slow non-Hermitian dynamics. The Riccati equation~\eqref{wx} for the population ratio $R_{\pm}$ maps the linear Schr\"odinger evolution onto effective nonlinear dynamics with instantaneous attractors $R_{\mathrm{ad}}(t)$ and $R_{\mathrm{nad}}(t)$ whose stability is controlled by the sign of $\operatorname{Im} E(t)$. In this picture, the transition from $R_{\rm ad}$ to
$R_{\rm nad}$ provides the dynamical signature of
irreversibility. The critical timescale $T_{\mathrm{cr}} \propto m\ln\beta$ thus provides a directly observable, purely forward-evolution manifestation of the PIR predictability horizon~\cite{PIR}, requiring no time-reversal echo to detect.

Our analytical predictions are consistent with the asymptotic selection rule~\cite{Uzdin2011,Milburn2015}
\begin{equation}\label{limit}
\lim_{T\to\infty}R_+(T)\,R_-(T)\to 1\,,
\end{equation}
which guarantees that a single eigenstate is eventually selected but does not specify which one or how fast. The explicit $T_{\mathrm{cr}}$ formula fills this gap. Our results also agree with the numerical observations of Ref.~\cite{Hassan2017a}: mode conversion becomes more efficient both for trajectories closer to the EP [Eq.~\eqref{res3}] and for larger loop contours [Eqs.~\eqref{res1}, \eqref{res2}, \eqref{res3}], as predicted by the geometry-dependent scaling of $\mathcal{G}$.

Looking ahead, the formula $T_{\mathrm{cr}} = \mathcal{G}\,\ln(1/|\Delta|)$ provides experimentalists with a concrete, testable prediction: for a given loop geometry, the critical timescale can be extracted from the onset of state conversion. In coupled-waveguide or fiber-loop platforms~\cite{Doppler2016,Yoon2018}, this translates to a critical propagation length that scales with the logarithm of the dynamic range. The precision-dependent regime ($T_{\mathrm{cr}} \propto m$) could be probed using programmable platforms with controllable bit depth. More broadly, the connection between $T_{\mathrm{cr}}$ and chiral state conversion opens the possibility of using the critical timescale as a quantitative tool in proposals that link non-Hermitian loops to quantum measurement~\cite{foa_torres_non-hermitian_2025}. Extensions to higher-dimensional Hamiltonians and more general parameter trajectories are natural next steps.

\textit{Author Contributions.--}
This work originated from discussions between L.E.F.F.T. and V.A. during V.A.'s visit to Santiago, motivated by open questions on non-adiabatic transitions and state conversion in non-Hermitian loops, including the non-encircling case. G.P. and later D.B.A. joined the project. V.A. and G.P. introduced the dynamical-bifurcation approach and developed the Riccati equation analysis; G.P. carried out the analytical derivations of the geometry-dependent growth factors. D.B.A. performed independent numerical simulations and worked on the connection with Berry's Stokes analysis and performed the simulations of Figs. 4 and 5. L.E.F.F.T. and D.B.A. identified the role of finite precision as the instability seed, leading to the formulation of precision-induced irreversibility (PIR). G.P. and V.A. wrote a first draft. L.E.F.F.T. wrote Section~VI and rewrote the introduction and conclusions; D.B.A. read and commented on the manuscript. All authors discussed the results and revised the manuscript.

\textit{Acknowledgments.--}
D.B.A. and L.E.F.F.T. thank Igor Gornyi, Alexander Mirlin and Ihor Poboiko for useful discussions. We thank M. V. Berry for correspondence and for bringing Ref.~\cite{Berry2025} to our attention. L.E.F.F.T. acknowledges financial support by ANID FONDECYT (Chile) through grant 1250751, The Abdus Salam International Centre for Theoretical Physics and the Simons Foundation. D.B.A. acknowledges the financial support of ANID/Subdirección de Capital Humano through Beca Doctorado Nacional Chile/21250325.
V.A. and G.P. acknowledge the support of the European Research Council (ERC) under the European Union's Horizon Europe programme (ERC StG “NASA”, Grant Agreement No. 101077954). 

\appendix \addcontentsline{toc}{chapter}{APPENDICES} \section{Non-adiabatic transitions as dynamical bifurcations}\label{app:db} Defining the ratio $R_{\pm}:=c_{\mp}/c_{\pm}$ and using Eq.~\eqref{system}, we find that they satisfy \begin{equation}\label{wx} \dot{R}_{\pm}=\pm2iE(s)R_{\pm}\pm h(s)(1+R_{\pm}^2) \end{equation} where \begin{equation} E(s)=\sqrt{1-z^2(s)}\,,\quad h(s)=-\frac{i\dot{z}(s)}{2E^2(s)} \end{equation} with $s=ft$ and initial conditions $R_{\pm}(0)=0$ corresponding to $\psi(0)=v_{\pm}(0)$. As explained in Ref.~\cite{Milburn2015}, for $f\ll1$ Eq.~\eqref{wx} can be analyzed using the theory of dynamical bifurcations~\cite{LebovitzSchaar1977,Mishchenko1980,Neishtadt1987,Neishtadt1988,BerglundSchneider1999,Diener1991,Glendinning1994}. The idea is to study the bifurcation diagram of Eq.~\eqref{wx} with $s$ treated as a parameter, exploiting the fact that $s$ varies slowly when $f\ll1$. The fixed points are identical for both $R_{\pm}(s)$ and, under the adiabatic condition $|h(s)/2E(s)|\ll1$, are approximately \begin{equation}\label{fp_app} R_{ad}(s)\approx\frac{ih(s)}{2E(s)},\quad R_{nad}(s)\approx \frac{2E(s)}{ih(s)} \end{equation} A stability analysis shows that for $R_+(t)$, $R_{\mathrm{ad}}(s)$ is stable if and only if $\Im E(s)>0$, while $R_{\mathrm{nad}}(s)$ is stable if and only if $\Im E(s)<0$. For $R_-(t)$ the situation is reversed. When $\Im E(s)=0$ neither fixed point is stable; we denote these values by $s_*$. When $s=ft$ varies slowly ($f\ll 1$), the solution behavior depends on whether $s$ crosses the bifurcation points $s_*$. If it does not, the solution tracks the stable fixed point arbitrarily closely~\cite{LebovitzSchaar1977,Mishchenko1980}. If $s$ crosses $s_*$, however, the stability is exchanged, and the solution continues to follow the now-unstable branch for a finite time before jumping to the stable equilibrium~\cite{Neishtadt1987,Neishtadt1988,BerglundSchneider1999}. As analyzed in Ref.~\cite{Milburn2015}, the linearized solution of Eq.~\eqref{wx} captures this behavior. One finds \begin{equation}\label{gen} R_{\pm}(t)=\pm e^{\pm W(t)}\int_0^th(\tau)e^{\mp W(\tau)}d\tau \end{equation} where \begin{equation} W(t)=2i\int_0^tE(\tau)d\tau \end{equation} Without loss of generality, we focus on $R_+(t)$. Considering times near $t_*$—which we assume corresponds to the point where $R_{ad}(t)$ becomes unstable and $R_{nad}(t)$ stable—we perform $N$ integrations by parts. Using properties of asymptotic series, Eq.~\eqref{gen} can be written as \begin{equation}\label{wwx} R_+(t)\approx\mathcal{R}^{(N)}_{ad}(t)-\mathcal{R}^{(N)}_{ad}(0)e^{W(t)}+\Delta(t) e^{\Psi(t)} \end{equation} where \begin{equation}\label{am} \mathcal{R}^{(N)}_{ad}(t)=\sum_{n=0}^{N-1}\left(\frac{i}{2E(t)}\frac{d}{dt}\right)^nR_{ad}(t) \end{equation} and \begin{equation} \Psi(t)=2i\int_{t_*}^tE(\tau)d\tau \end{equation} The first two terms in Eq.~\eqref{wwx} arise from the endpoints of the integration interval, while the third is the remainder, with $\Delta(t)$ a time-dependent function. Under general assumptions discussed in Ref.~\cite{Milburn2015}, the remainder in the large-$T$ limit can be approximated as $\Delta(t)\approx\Theta(t-t_*)\Delta$, where $\Delta$ is a time-independent function of the system parameters. The Heaviside function reflects the fact that for $t<t_*$, when $R_{\mathrm{ad}}(t)$ is stable, the last term is negligible. We focus on times close to the end of the period, although the arguments hold for general $t$. For most cases considered in the main text, the final term in Eq.~\eqref{wwx} ultimately drives the deviation from $R_{ad}(t)$, producing the non-adiabatic transition. The solution tracks the unstable fixed point for a finite duration before jumping to the stable one, giving rise to a delay time $t_+$. This delay is determined by the last term in Eq.~\eqref{wwx} and is defined by \begin{equation}\label{dt} |\Delta e^{\Psi(t_+)}|=1 . \end{equation} For the purposes of this work we seek the relation between the system parameters that allows sufficient time for the transition to $R_{nad}(t)$ to occur near the end of the period. To this end, we use Eq.~\eqref{dt} with $t_+$ replaced by $T$, which we denote by $T_{cr}$. In other words, $T_{cr}$ determines the degree of slowness required for the system to reach the preferred state. As seen from Eq.~\eqref{dt}, $T_{cr}$	
  depends directly on 
$\Delta$ which represents the seed of the instability. Its origin for the EP-encircling and non-EP-encircling trajectories will be studied in detail in Sec. \ref{sec:seed}. 
  \section{Non-encircling loops for $T<T_{cr}$}\label{app:accuracy} 
\begin{figure}[h] \includegraphics[width=\linewidth]{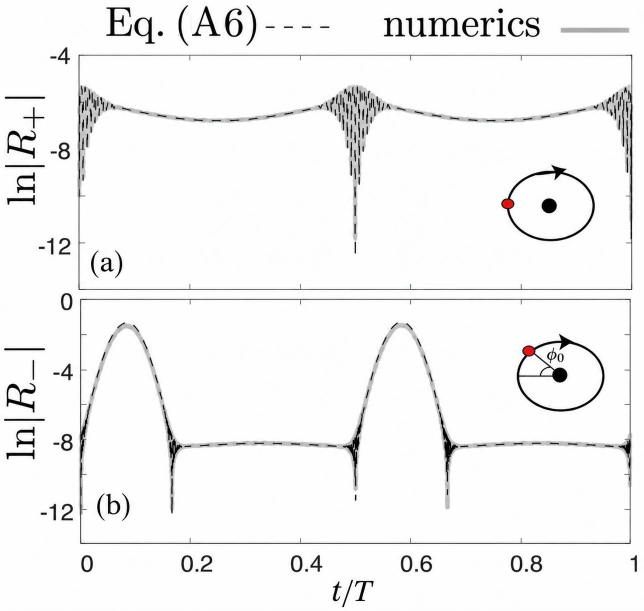} \caption{(a) Time evolution of $\ln|R_+(t)|$ for the symmetric loop with $r=0.5$ and $T=500$. (b) Time evolution of $\ln|R_-(t)|$ for the symmetric loop with $r=0.3$, $\phi_0 = \pi/3$ and $T=2000$. In both panels, the grey solid curve is the numerical solution and the dashed curves represent the analytical approximation obtained from \eqref{wwx} with the final term neglected and the adiabatic expansion
$\mathcal{R}_{\rm ad}^{(N)}$ of Eq.~(A7) truncated at $n=0$.}
  \label{fig:Figure6}
  \end{figure}

For $T<T_{\mathrm{cr}}$ we find that the last term in Eq.~\eqref{wwx} remains small and $R_{\pm}(t)$ is well approximated by the first two terms, i.e., the endpoint contributions. This is illustrated in Fig.~\ref{fig:Figure6}(a,b) where we compare the numerical solution with the analytical one, obtained from \eqref{wwx} with the last term neglected.
  
  For $T<T_{\mathrm{cr}}$, the growth and decay are governed by $\Re W(t)$, which involves an integral over the full period. For non-EP-encircling trajectories $\Re W(t\to T)\to0$, $\forall T$. Thus as long as the solution is well described by the first two terms of Eq.~\eqref{wwx}, it remains close to $R_{\mathrm{ad}}(T)$ at the end of the evolution and the final state coincides with the initial one. Especially for the symmetric loop $(g_0=0,\phi_0=0)$, the system evolves adiabatically throughout the cycle. This is a direct consequence of this trajectory's geometry. Using \eqref{en} we find 
  \begin{equation}
    W(t)=2it+\frac{r^2}{2f}(e^{-2ift}-1)
  \end{equation}
  thus the following inequality holds
  \begin{equation} \begin{split} |\mathcal{R}^{(N)}_{ad}(t)-\mathcal{R}^{(N)}_{ad}(0)e^{W(t)}|\leq |\mathcal{R}^{(N)}_{ad}(t)|+|\mathcal{R}^{(N)}_{ad}(0)|\ll1 . \end{split} \end{equation}
  According to the above, in order to capture the observed deviation from $R_{ad}(t)$ for even larger periods, the last term in \eqref{wwx} must be taken into account.

   \section{Case study: Encircling the EP}\label{app:EPenc} Suppose the Hamiltonian \eqref{sy} and consider a circular trajectory enclosing the EP at $+1$, setting $g_0=1$ and $\phi_0=\pi+\theta_0$ with $\theta_0\in[0,\pi)$ to facilitate comparison with Ref.~\cite{Berry2011}. We focus on $R_-(t)$, for which $R_{\mathrm{ad}}(t)$ becomes unstable near the end of the evolution. For $r\ll1$, \begin{equation}\label{approxberry} E(t)\approx-i\sqrt{2r}e^{-\frac{i}{2}(ft+\theta_0)},\quad h(t)\approx -\frac{if}{4} \end{equation}
   Let us note that Eq.~\eqref{wx} for the above approximate expressions is the Riccati equation of the model in Ref.~\cite{Berry2011}, valid for arbitrary $r$. The corresponding solution is given in \eqref{gen} with \begin{equation}\label{stokes} \begin{split} &W(t)=Z(t)-Z(0)\\ &Z(t)=\frac{4i\sqrt{2r}}{f}e^{-\frac{i}{2}(ft+\theta_0)} \end{split} \end{equation} As discussed in Ref.~\cite{Milburn2015} (see also references therein for the exponential integral function) the remainder in Eq.~\eqref{wwx}, evaluated for $Z(t)$ given in Eq.~\eqref{stokes}, generates a Stokes contribution $2\pi i$ each time $\arg Z(t)$ crosses the negative real axis, in this case at $ft_*=\pi-\theta_0$ which is also the time where $R_{ad}(t)$ becomes unstable and $R_{nad}(t)$ stable. Taking into account the above we write the linearized solution around $t_*$ as  \begin{equation}\label{solst} 
  \begin{split} R_-(t)&=\mathcal{R}^{(N)}_{ad}(t)-\mathcal{R}^{(N)}_{ad}(0)e^{-W(t)}\\ &+\left(-i\pi-\Delta e^{Z(t_*)}\right)\Theta(t-t_*)e^{-Z(t)} \end{split} 
  \end{equation}
  where $R_{\mathrm{ad}}(t)=\frac{if}{8\sqrt{2r}}e^{i(ft+\theta_0)/2}$. Notice that for the case $\theta_0=\pi$ there is no Stokes contribution as $Z(t)$ does not cross the negative real axis. 
  
  The special case $\theta_0 = 0$ corresponds to the scenario discussed in Sec.~\ref{Encircling}. For periods $T < T_{cr}$, the last term in $\eqref{solst}$ can be neglected. In this regime, both $W(T)$ and $Z(T)$ are purely imaginary, leading to $|R_-(T)| \approx \pi$. This constant value arises from the Stokes contribution and accounts for the first plateau observed in Fig.~\ref{fig:Figure2}(f), which is independent of $T$ and $r$, and notably larger than $|R_{ad}(t)|$. We note, however, a discrepancy between the Stokes contribution obtained here and the numerical value of \( |R_-(T)| \) in the first plateau. In Ref.~\cite{Berry2011}, the authors employed Bessel asymptotics to evaluate \( |R_-(T)| \), obtaining exactly the value \( 2 \). We attribute this difference primarily to the distinct asymptotic expansions used.
  
For $\theta_0 \in (0,\pi)$, numerical results indicate that no critical period $T_{cr}$ can be identified. In this case, the notion of a transition scale breaks down, as $|R_-(T)|$ has already reached the nonadiabatic value $|R_{nad}|$ for sufficiently small $T$. This behavior is consistent with the condition $\Re{W(T)} < 0$ in this parameter regime.
  \newpage
\section{Coefficient magnitudes across the transition}\label{app:coeffs} 

\begin{figure}[h!] \includegraphics[width=1\linewidth]{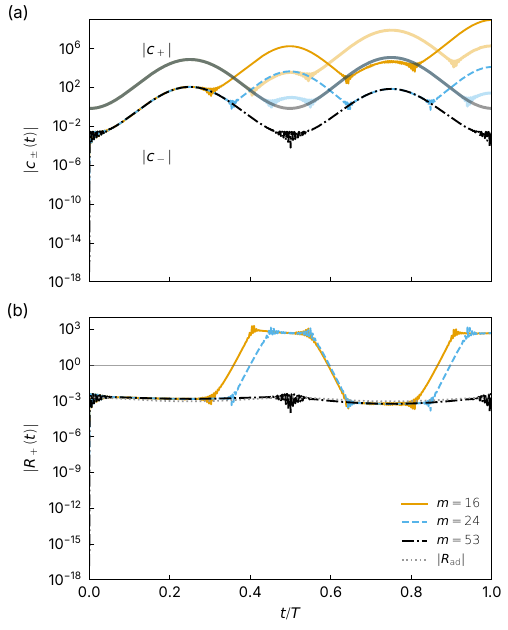} \caption{(a)~Coefficient magnitudes $|c_\pm(t)|$ for the symmetric loop $z(t)=-re^{-ift}$ with $r=0.5$ and $T=600$, initialized in $v_+(0)$, at $m=16$, $24$, and $53$ mantissa bits (thick faded curves: $|c_+|$; thin curves: $|c_-|$). (b)~The population ratio $|R_+(t)|$ for the same runs; the dotted curve is the adiabatic fixed point $|R_{\mathrm{ad}}(t)|$ and the horizontal line marks $|R_+|=1$. The early-time transient starts at the rounding floor $\beta^{-m}$ of each run.}\label{fig:coeffs}\end{figure}

A natural question is whether the critical time could instead be attributed to the coefficients $c_{\pm}(t)$ becoming too small to represent, an underflow effect of the dissipative terms, rather than to the amplification of a rounding-scale seed. Figure~\ref{fig:coeffs} rules this out for the symmetric loop ($r=0.5$, $T=600$, $\phi_0=0$), initialized in $v_+(0)$ and integrated with the same fourth-order Runge--Kutta scheme and time step $dt=10^{-2}$ as in the main text, at $m=16$, $24$, and $53$ mantissa bits. Three features are relevant. First, $|c_+(t)|$ never drops below its initial value $\simeq 0.71$ at any precision,  and the three envelopes coincide to $1.5\times10^{-4}$ in $\log_{10}|c_+|$ until the first non-adiabatic transition; the visible splitting afterwards reflects the different final states, not an accumulation of error. Second, the smallest nonzero values of $|c_-|$---$9.0\times10^{-6}$, $1.3\times10^{-8}$, and $1.0\times10^{-17}$---are $\simeq\beta^{-m}$: the rounding floor itself, appearing in the initially empty mode during the first steps of the evolution, i.e., the seed of Sec.~\ref{sec:seed} made directly visible. Third, with identical equations, integrator, and time step, the population ratio crosses unity at $t/T=0.36$ and $0.40$ for $m=16$ and $24$, while for $m=53$ it never leaves the adiabatic branch ($|R_+(T)|=4.8\times10^{2}$, $5.0\times10^{2}$, and $1.8\times10^{-4}$, respectively): $T=600$ lies between $T_{\mathrm{cr}}(m=24)$ and $T_{\mathrm{cr}}(m=53)$, consistent with Eq.~\eqref{Tcr_precision}. In arbitrary-precision arithmetic the mantissa length and the exponent range are set independently; no underflow occurs anywhere in these runs. The transition is carried by the ratio $R_+$, not by the magnitude of the wavefunction coefficients.

%
\end{document}